%% file: m3j.tex
\documentclass[aps,prl,showpacs,superscriptaddress,twocolumn,amsfonts,floatfix]{revtex4}
\usepackage{graphicx}
\usepackage{dcolumn}
\usepackage{bm}
\usepackage{epsfig}
\usepackage{amssymb,amsmath}
\usepackage{slashed}
\suppressfloats[!]
\usepackage{multirow}

\newcommand{\pT}{p_T}
\newcommand{\Mtj}{M_{\text{3jet}}}
\newcommand{\as}{\alpha_s}
\newcommand{\asmz}{\alpha_s(M_Z)}
\newcommand{\sherpa}{{\sc sherpa}}
\newcommand{\pythia}{{\sc pythia}}
\newcommand{\ord}{{\cal O}}
\newcommand{\ppbar}{p{\bar{p}}}
\newcommand{\Rcone}{{\cal R}_{\text{cone}}}

\begin{document}

\hspace{5.2in}\mbox{FERMILAB-PUB-11-173-E}

\title{\boldmath Measurement of three-jet differential cross sections 
   $d\sigma_{\text{3jet}} / dM_{\text{3jet}}$ in $p\bar{p}$ collisions 
   at $\sqrt{s}=1.96$ TeV}

\input author_list.tex

\date{April 14, 2011}

\begin{abstract}\noindent
We present the first measurement of the inclusive three-jet differential 
cross section as a function of the invariant mass of the three jets 
with the largest transverse momenta in an event in
$p\bar{p}$ collisions at $\sqrt{s}=1.96\, \mathrm{TeV}$.
The measurement is made in different rapidity regions
and for different jet transverse momentum requirements
and is based on a data set corresponding to an integrated luminosity
of $0.7\, \mathrm{fb}^{-1}$
collected with the D0 detector at the Fermilab Tevatron Collider.
The results are used to test the three-jet matrix elements 
in perturbative QCD calculations at next-to-leading order 
in the strong coupling constant.
The data allow discrimination between parametrizations of 
the parton distribution functions of the proton.
\end{abstract}

\maketitle


\begin{figure*}[t]
\centering
\includegraphics[scale=1.03]{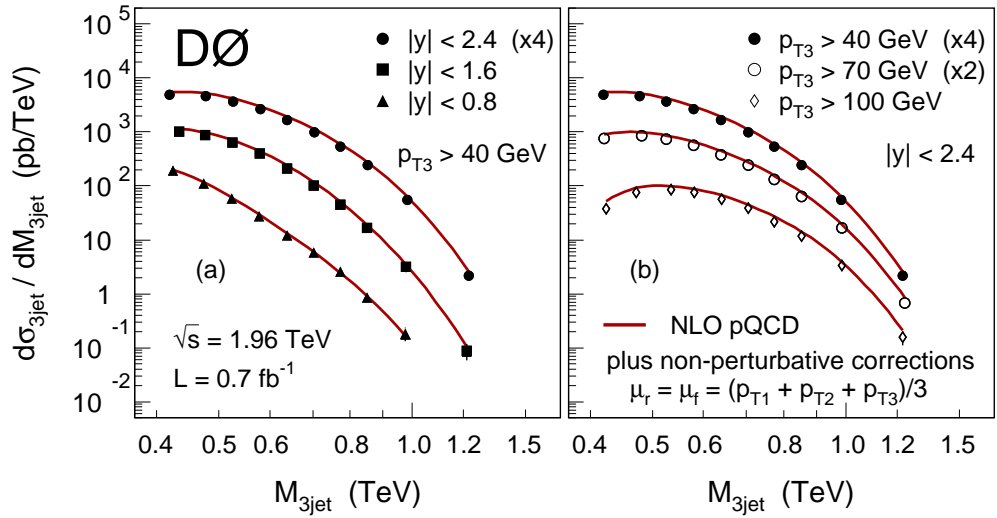}
\caption{(Color online.)
The differential cross section $d\sigma_{\text{3jet}} / d\Mtj$
(a) in different rapidity regions and 
(b) for different $p_{T3}$ requirements.
The solid lines represent the NLO pQCD matrix element calculations using MSTW2008NLO PDFs
and $\as(M_Z)=0.1202$ which are corrected for non-perturbative effects.
\label{fig:fig1}}
\end{figure*}

The production cross section for jets with large transverse momenta ($\pT$)
with respect to the beam axis in hadron-hadron collisions
is predicted by perturbative QCD (pQCD)
and is sensitive to the strong coupling constant ($\as$) and the
parton distribution functions (PDFs) of the hadrons.
Deviations from the pQCD predictions may indicate the presence
of physics processes not included in the standard model.
Recent measurements of inclusive jet and dijet production in
$\ppbar$ collisions at a center-of-mass energy of 
$\sqrt{s}=1.96$\,TeV~\cite{Abulencia:2007ez,Aaltonen:2008eq,Aaltonen:2008dn,:2008hua,:2009mh,Abazov:2010fr} 
have been used to determine $\as$~\cite{Abazov:2009nc} 
and the proton PDFs~\cite{Martin:2009iq,Lai:2010vv,Ball:2011mu}
and to set limits on a number of models of physics beyond 
the standard model~\cite{Aaltonen:2008dn,:2009mh}.
This demonstrates the success of pQCD in describing observables
which are directly sensitive to the matrix elements of $\ord(\as^2)$.
Testing pQCD at higher orders of $\as$ requires measuring
cross sections for higher jet multiplicities.

The three-jet cross section is directly sensitive 
to the pQCD matrix elements of $\ord(\as^3)$, 
and therefore has a higher sensitivity to $\as$
as compared to inclusive jet and dijet cross sections,
while having a similar sensitivity to the PDFs.
Since pQCD calculations are available to next-to-leading order (NLO) 
in $\as$~\cite{GieleKilgore,Kilgore:1996sq,Nagy:2003tz,Nagy:2001fj},
the three-jet cross section can be used for precision phenomenology
such as simultaneous determinations of $\as$ and PDFs from experimental data.
In such QCD analyses~\cite{Martin:2009iq,Alekhin:2009ni}, 
the information from three-jet cross sections can supplement that
from inclusive jet and dijet cross sections,
partially decorrelating the results for $\as$ and the PDFs.

In this Letter, we present the first measurement of 
the inclusive three-jet differential cross section, 
$d\sigma_{\text{3jet}} / d\Mtj$,
in $\ppbar$ collisions at $\sqrt{s}=1.96$\,TeV,
as a function of the invariant mass ($\Mtj$) 
of the three highest-$\pT$ jets in each event.
The data sample, collected with the D0 detector~\cite{d0det}  
during 2004--2005 in Run~II of the Fermilab Tevatron Collider, 
corresponds to an integrated luminosity of $0.7\,$fb$^{-1}$.
In the experiment and in the theoretical calculations used in this analysis, 
jets are defined by the Run~II midpoint cone jet algorithm~\cite{run2cone} 
with a cone of radius $\Rcone =0.7$ in rapidity $y$ 
and azimuthal angle $\phi$.
Rapidity is related to the polar scattering angle $\theta$
with respect to the proton beam axis by 
$y=\frac{1}{2} \ln \left[ (1+\beta\cos\theta)/(1-\beta \cos \theta) \right]$,
where $\beta$ is defined as the ratio between momentum and energy
($\beta=|\vec{p}| / E$).
The inclusive three-jet event sample consists of all events
with three or more jets which pass given $p_T$ and $|y|$ requirements.
The $\Mtj$ dependence of the inclusive three-jet cross section 
is measured for five scenarios with different jet $\pT$ requirements 
and in different regions of jet rapidity.
Jets are ordered in descending $\pT$ and the $\pT$ requirements are 
$p_{T1} > 150$\,GeV and $p_{T3} > 40$\,GeV
(with no further requirement for $p_{T2}$).
The rapidities of the three leading $\pT$ jets are restricted to
$|y| < 0.8$, $|y| < 1.6$, or $|y| < 2.4$, in three different measurements.
Two additional measurements are made for $p_{T3} > 70$\,GeV 
and $p_{T3} > 100$\,GeV, both requiring $|y| < 2.4$.
For jets defined by the cone radius $\Rcone$ and for a given $p_{T3}$ 
requirement, the relative $p_T$ between two jets ($k_\perp$)
could be as low as $k_\perp \approx \Rcone \cdot p_{T3}$,
which introduces an additional, softer scale in the process
(since $k_\perp < p_{T3}$  for $\Rcone= 0.7$).
The phase space with $k_\perp$ below the $p_{T3}$ requirement
can be avoided by an additional requirement on the
angular separation of the three leading $\pT$ jets.
In all scenarios, all pairs of the three leading $\pT$ jets are required 
to be separated by 
$\Delta R = \sqrt{(\Delta y)^2 + (\Delta \phi)^2}> 1.4 \; (= 2 \cdot \Rcone)$.
With this separation requirement, the smallest accessible $k_\perp$
of the jets is always above $p_{T3}$.
Furthermore, this separation requirement also reduces the phase space 
in which pairs of the three leading $p_T$ jets are subject to 
the overlap treatment in the cone jet algorithm~\cite{run2cone}.
Since the overlap treatment can strongly depend on details
of the energy distributions in the overlap area,
this region of phase space may not be well modeled by pQCD
calculations at lower orders.
In the remaining analysis phase space, NLO pQCD calculations
are not affected by the Run II cone algorithm's 
infrared sensitivity~\cite{Salam:2007xv}.
The data are corrected for instrumental effects and are presented 
at the ``particle level,'' which includes all stable particles 
as defined in Ref.~\cite{Buttar:2008jx}.

\begin{figure*}[t]
\centering
\includegraphics[scale=1]{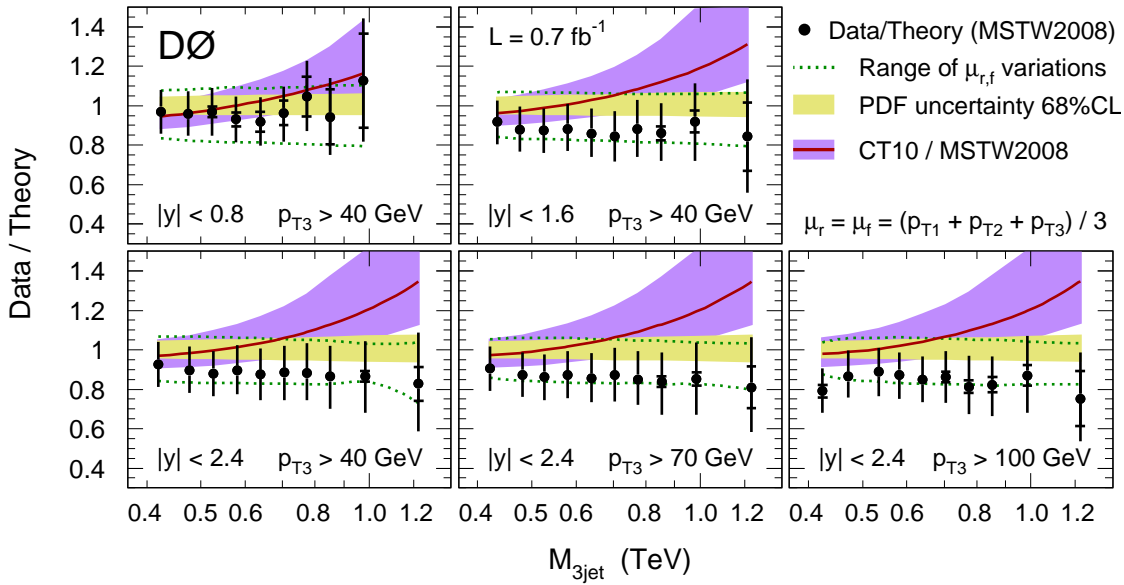}
\caption{(Color online.)
Ratios of the differential cross sections
$d\sigma_{\text{3jet}} / d\Mtj$
measured in different rapidity regions and for different 
$p_{T3}$ requirements
and the pQCD predictions for different PDFs.
The inner uncertainty bars indicate the statistical uncertainties,
and the total uncertainty bars display the quadratic sum of 
the statistical and systematic uncertainties.
The ranges of renormalization and factorization scale variations 
(as specified in the text) are indicated by the dotted lines,
while the PDF uncertainties are indicated by the shaded bands.
\label{fig:fig2}}
\end{figure*}


A detailed description of the D0 detector can be found in
Ref.~\cite{d0det}.
The event selection, jet reconstruction, and jet energy and 
momentum correction in this measurement follow closely those 
used in our recent inclusive jet and dijet 
measurements~\cite{:2008hua,:2009mh,Abazov:2010fr}.
Jets are reconstructed in the finely segmented
D0 liquid-argon/uranium calorimeter which covers most of the solid angle
for polar angles of
$1.7^\circ \lesssim \theta \lesssim 178.3^\circ$~\cite{d0det}.
For this measurement, events are triggered by the jet with highest $p_T$.
Trigger efficiencies are studied by comparing the three-jet cross section
in data sets obtained by inclusive jet triggers with different 
$\pT$ thresholds 
in regions where the trigger with lower threshold is fully efficient. 
The trigger with lowest $\pT$ threshold is shown to be fully efficient
by studying an event sample obtained independently with a muon trigger.
In each $\Mtj$ bin, events are taken from a single trigger
which is chosen such that its efficiency is above 99\%.

The position of the $p\bar{p}$ interaction is determined from 
the tracks reconstructed in the silicon detector and 
scintillating fiber tracker located inside a $2\,\text{T}$ 
solenoidal magnet~\cite{d0det}.
The position is required to be within $50$\,cm of the detector center
in the coordinate along the beam axis, 
with at least three tracks pointing to it.
These requirements discard (7.1--8.6)\% of the events. 
Contributions from cosmic ray events are suppressed by 
requiring the missing transverse energy ($\slashed{E}_T$)
in an event to be $\slashed{E}_{T}< 0.5 \cdot p_{T1}$.
This requirement is applied before the jet four-momenta are corrected,
and its efficiency for signal is found to be  $>99.5\%$~\cite{:2008hua}.
Requirements on characteristics of calorimeter shower shapes are
used to suppress the remaining background due to electrons, photons, 
and detector noise that would otherwise mimic jets. 
The efficiency for the shower shape requirements is above $97.5\%$, 
and the fraction of background events is below $0.1$\% for all $\Mtj$.

\begin{figure*}[t]
\centering
\includegraphics[scale=1]{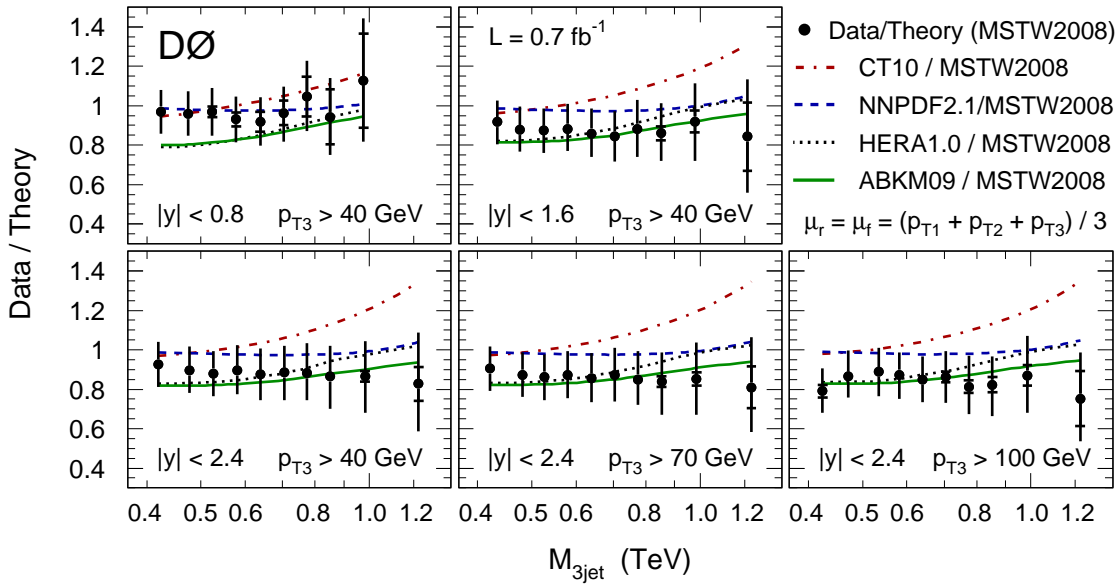}
\caption{(Color online.)
Ratios of the differential cross sections
$d\sigma_{\text{3jet}} / d\Mtj$
measured in different rapidity regions and for different 
$p_{T3}$ requirements and the pQCD predictions for different PDFs.
The inner uncertainty bars indicate the statistical uncertainties,
and the total uncertainty bars display the quadratic sum of 
the statistical and systematic uncertainties.
\label{fig:fig3}}
\end{figure*}

The jet four-momenta reconstructed from calorimeter energy depositions
are then corrected, on average, for the response of the calorimeter, 
the net energy flow through the jet cone,
additional energy from previous and consecutive beam crossings, and 
multiple $p\bar{p}$ interactions in the same event, but not for 
muons and neutrinos.
The absolute energy calibration is determined from 
$Z \rightarrow ee$ events and the 
$p_T$ imbalance in $\gamma$ + jet events in the region $|y| < 0.4$.
The extension to larger rapidities is derived from dijet events
using a similar data-driven method.
In addition, corrections are applied which take
into account the difference in calorimeter response due to the
difference in the fractional contributions of quark- and 
gluon-initiated jets in the dijet and the $\gamma$ + jet event samples.
These corrections of the order (2--4)\% are determined using simulated jets 
produced with the \pythia\ event generator~\cite{pythia} that have been 
passed through a {\sc geant}-based detector simulation~\cite{geant}.
The total corrections for the jet four-momenta vary between 50\% and 
20\% for jet $\pT$ between 50 and 400\,GeV.
These corrections adjust the reconstructed jet energy to the
energy corresponding to the stable particles that entered the
calorimeter except for muons and neutrinos, which are accounted for
later by a separate correction.
An additional correction is applied for systematic shifts in $|y|$ 
due to detector effects~\cite{:2008hua}.
The three-jet invariant mass is then computed from the corrected 
jet four-momenta of the three highest-$p_T$ jets in the event.


The differential cross sections $d\sigma_{\text{3jet}} / d\Mtj$ 
are corrected for experimental effects~\cite{Hubacek:2010zza}.
Particle-level jets from events generated with \sherpa~\cite{sherpa} 
with MSTW2008LO PDFs~\cite{Martin:2009iq} are processed by a fast simulation 
of the D0 detector response.
The simulation is based on parametrizations of 
resolution effects in $p_T$, the polar and azimuthal angles of jets,
jet reconstruction efficiencies, and misidentification of the event vertex,
which are determined either from data or from a detailed simulation 
of the D0 detector using {\sc geant}.
The $\pT$ resolution for jets is about 15\% at 40 GeV, decreasing 
to less than 10\% at 400 GeV.
The generated events are reweighted to match the $\Mtj$, $p_T$, 
and $|y|$ distributions in data.
To minimize migrations between $\Mtj$ bins due to resolution effects,
we use the simulation to obtain a rescaling function in reconstructed $\Mtj$
that optimizes the correlation between the reconstructed and true values.
The bin sizes in the $\Mtj$ distributions are chosen to be 
approximately twice the $\Mtj$ resolution.
The bin purity after $\Mtj$ rescaling, defined as the fraction
of all reconstructed events that were generated in the same
bin, is above 40\% for all bins.
We then use the simulation to determine $\Mtj$ bin correction factors
for instrumental effects
for the differential cross sections in the five different scenarios.
These also include corrections for the energies of unreconstructed
muons and neutrinos inside the jets.
The total correction factors for the differential cross sections 
vary from about 1.0 at $\Mtj = 0.4$\,TeV to 1.1 at 1.1\,TeV 
for $|y|<0.8$ and between 0.89 at $\Mtj =0.4$\,TeV to 0.96 at 1.1\,TeV 
for $|y|<2.4$.
The dependence of the correction factors on the reweighting function
is taken into account as an uncertainty.
The corrected differential cross section in each scenario is presented 
at the ``particle level'' as defined in Ref.~\cite{Buttar:2008jx}.

In total, 65 independent sources of experimental systematic uncertainties
are identified, mostly related to jet energy and jet $p_T$ resolution.
The effects of each source are taken as fully correlated between 
all data points.
The dominant uncertainties for the differential cross sections are due 
to the jet energy calibration [(10--30)\%], 
the luminosity uncertainty (6.1\%), and
the jet $p_T$ resolution [(1--5)\%].
Smaller contributions come from the uncertainties in systematic shifts 
in $y$ (3\%), reweighting of the generated events (2.5\%),
trigger efficiency uncertainties (2\%), 
and from the jet $\theta$ resolution (1\%).
All other sources are negligible.
The systematic uncertainties are never larger than 30\%,
and for $\Mtj<0.9$\,TeV, they are between 11\% and 20\%.

\input{m3j-table}


The results for the differential cross sections 
for different rapidity and $p_{T3}$ requirements
are given in Table~\ref{tab:data} and displayed in Fig.~\ref{fig:fig1}.
A detailed documentation of the results, including the individual 
contributions from all 65 sources of correlated uncertainties
is provided in the supplemental material~\cite{supplement}.  
The quoted central values of $\Mtj$ at which the data points are 
presented are the locations where the bin averages
have the same value as the differential cross section~\cite{Lafferty:1994cj}, 
as determined using smooth parametrizations of the data.
The data are compared to theory predictions which have been obtained from 
NLO pQCD calculations with non-perturbative 
corrections applied.
The non-perturbative corrections are determined using \pythia\
with ``tune DW''~\cite{Albrow:2006rt}.
They are defined as the combination of the corrections due to hadronization
and underlying event and vary between $-10\%$ and $+2\%$
(given in Table~\ref{tab:data}).
Using different \pythia\ settings (A, BW, Z1, Perugia soft, Perugia hard tunes)
affects the individual corrections by less than half of their sizes
and the total corrections by less than 5\%.
The NLO results are computed using {\sc fastnlo}~\cite{Kluge:2006xs}
based on {\sc nlojet++}~\cite{Nagy:2003tz,Nagy:2001fj}
with MSTW2008NLO PDFs~\cite{Martin:2009iq}
and the corresponding value of $\as(M_Z)=0.1202$.
The central choice $\mu_0$ for the renormalization and factorization scales 
is the average $\pT$ of the three leading $\pT$ jets 
$ \mu_r = \mu_f = \mu_0 = (p_{T1}+p_{T2}+p_{T3})/3$.
For a direct comparison of the theoretical predictions with data, the
ratio of data and theory is displayed by the markers in Fig.~\ref{fig:fig2} 
for all five scenarios.
The effects of independent variations of renormalization 
and factorization scales between $\mu_0 / 2 $ and $2\mu_0$ 
are displayed by the dotted lines.
These variations affect the predicted cross sections between $+$(5--10)\% 
and $-$(15--20)\%.  

The MSTW2008NLO PDF uncertainties (corresponding to the 68\% C.L.) 
are shown by the light band.
The ratios of data and theory are almost constant, 
with only a small dependence on $\Mtj$ and the $|y|$ and $p_{T3}$ requirements.
The central data values are below the central theory predictions,
by approximately (4--15)\% in the different scenarios,
slightly increasing with $|y|$ and with $p_{T3}$.
In all cases, the data lie inside the range covered by the scale variation. 

In addition to the MSTW2008NLO PDFs, Fig.~\ref{fig:fig2} shows
also predictions for CT10 PDFs~\cite{Lai:2010vv} and the corresponding value 
of $\as(M_Z)=0.118$, normalized by the predictions for MSTW2008NLO
and represented by the solid lines.
To compare the CT10 PDF uncertainties (which have been published 
at the 90\% C.L.) with the experimental uncertainties (corresponding to 
one standard deviation), the former have been scaled by a factor 
of 1/1.645~\cite{Ball:2008by}.
The resulting 68\% C.L.\ uncertainties are displayed around the
CT10 central values by the dark band.
The CT10 PDFs predict a different shape for the $\Mtj$ dependence
of the cross section.
For $\Mtj < 0.6\,$TeV, the central results for CT10 PDFs 
agree with those for MSTW2008NLO, while the CT10 predictions 
at $\Mtj = 1.2\,$TeV are up to 30\% higher.
These discrepancies at highest $\Mtj$
are larger than the combined 68\% C.L.\ uncertainty bands 
of the CT10 and MSTW2008NLO PDFs.

Calculations for additional PDFs are compared to the data in
Fig.~\ref{fig:fig3}.
These are the PDF parametrizations 
NNPDFv2.1~\cite{Ball:2011mu} ($\asmz = 0.119$), 
ABKM09NLO~\cite{Alekhin:2009ni} ($\asmz = 0.1179$),
and HERAPDFv1.0~\cite{:2009wt} ($\asmz = 0.1176$).
The results for NNPDFv2.1 agree everywhere within $\pm$4\% with those
from MSTW2008NLO.
The cross sections predicted for HERAPDFv1.0 are (15--20)\% below those 
for CT10 everywhere and their $\Mtj$ distributions have a similar shape.
The $\Mtj$ dependence of the calculations for the ABKM09NLO PDFs
is between the shapes of MSTW2008NLO/NNPDFv2.1 and CT10/HERAPDFv1.0.
At low $\Mtj$, the predictions for ABKM09NLO agree with those
for HERAPDFv1.0, while at higher $\Mtj$,
they predict the smallest cross sections of all PDFs under study.

The level of agreement between theory and data
can not be directly judged from the comparisons in Figs.~\ref{fig:fig2}
and~\ref{fig:fig3}, but requires taking into account correlations
of experimental uncertainties.
While some experimental uncertainties (like the luminosity uncertainty)
allow to shift the data points coherently up or down,
other uncertainty sources (such as the jet energy calibration),
have $\Mtj$-dependent effects
which also allow changes to shapes of the data distributions.
To quantify the significance of the differences
between theory and data in normalization and shape 
as observed in Figs.~\ref{fig:fig2} and~\ref{fig:fig3},
a $\chi^2$ is computed.
The $\chi^2$ definition takes into account  
all experimental uncertainties and their correlations,
as well as uncertainties in the hadronization and
underlying event corrections.
The latter two uncertainties are assumed to be half the size of the 
individual corrections,
to be independent of each other, and each to be fully correlated over $\Mtj$.
Correlations between the statistical uncertainties are ignored,
since the overlap of the data for the different scenarios is not large.
PDF uncertainties are not taken into account in the $\chi^2$ calculations.
Otherwise, a theoretical prediction
affected by large PDF uncertainties and in poor
agreement with data may get a smaller $\chi^2$ than
a prediction with better agreement with data but small
PDF uncertainties.
Therefore, since the PDF uncertainties are defined differently
for different PDF parametrizations~\cite{DeRoeck:2011na},
the $\chi^2$ values would no longer be suited 
to benchmark different PDF parametrizations.
This means that the $\chi^2$ values presented here are only a measure
of the agreement of the {\em central} PDF fit results with the 
measured three-jet cross sections.

\begin{figure*}[t]
\centering
\includegraphics[scale=1]{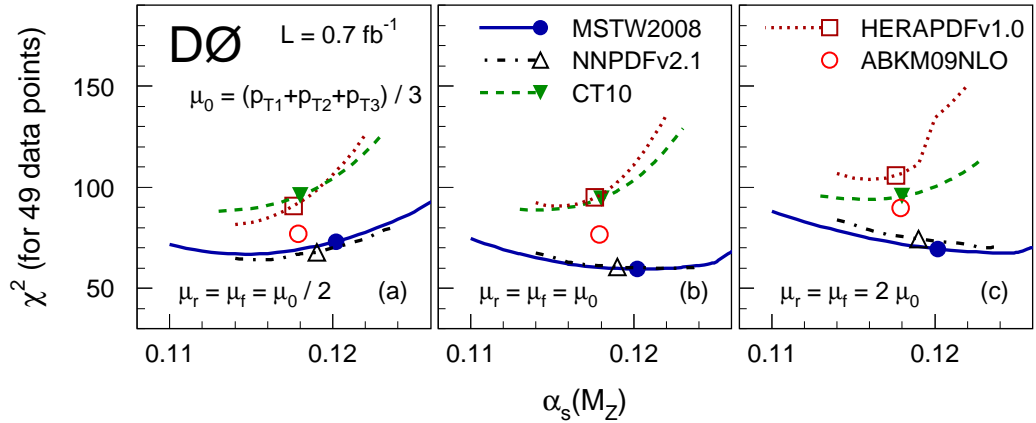}
\caption{(Color online.)
The $\chi^2$ values between theory and data, 
as a function of the value of $\asmz$ used in the matrix elements
and PDFs. 
The results are shown for different PDF parametrizations
and for different choices of the renormalization and 
factorization scales.
The positions of the central $\asmz$ values in the different PDF sets
are indicated by the markers.
\label{fig:fig4}}
\end{figure*}

The theory results, and therefore the $\chi^2$ values, depend on 
the choices of $\asmz$ and the scales $\mu_r$ and $\mu_f$ 
used in the computations of the NLO matrix elements
and on the chosen PDF parametrization.
The latter also depend implicitly on the value of $\asmz$.
All these dependencies are shown in Fig.~\ref{fig:fig4},
where the $\chi^2$ results are displayed
as a function of $\asmz$ used in the 
NLO matrix elements and PDFs,
for three alternative scale choices
$\mu_r=\mu_f= \mu_0/2$ (a), $\mu_0$ (b), and $2\mu_0$ (c).
The results for the central $\asmz$ choices for the different
PDF sets are also indicated.
For $\asmz$ values close to the world average of 
$0.1184\pm0.0007$~\cite{Bethke:2009jm},
for all PDF sets, with the exception of HERAPDFv1.0,
the lowest $\chi^2$ is obtained
for the central scale choice $\mu_r=\mu_f= \mu_0$.
Table~\ref{tab:table0} gives a summary of the $\chi^2$ values,
obtained using different PDFs with their default $\asmz$ values
for the central scale $\mu_r = \mu_f =\mu_0$,
as well as the minimum $\chi^2$ for all scale and $\asmz$ choices.
From all PDFs, the largest $\chi^2$ values are obtained for 
CT10 and HERAPDFv1.0 PDFs, independent of the scale and $\asmz$ choices.
These are always above $\chi^2 \ge 81.7$ for all 49 data points.
For ABKM09NLO PDFs, which are available only for 
a single value of $\asmz = 0.1179$,
the smallest $\chi^2$ is $76.5$, obtained for $\mu_r=\mu_f= \mu_0$.
The best overall agreement, corresponding to the lowest $\chi^2$ values,
is obtained for MSTW2008NLO 
for the central scale choice $\mu_r=\mu_f= \mu_0$ and $\asmz=0.121$
with $\chi^2 = 59.5$.
Very close to these are the results for NNPDFv2.1
for which the lowest $\chi^2$ is $59.9$ for 
$\mu_r=\mu_f= \mu_0$ and $\asmz =0.123$.
The large $\chi^2$ differences between the different PDF sets
demonstrate the PDF sensitivity of the three-jet cross section data.

\begin{table}
\caption{\label{tab:table0}
$\chi^2$ values between data and theory for different PDF parametrizations
in the order of decreasing $\chi^2$, for all 49 data points.}
\begin{ruledtabular}
\begin{tabular}{lccr}
PDF set & Default  & $\chi^2$ at $\mu_{r}=\mu_{f}=\mu_0$ & $\chi^2_{\rm minimum}$ \\
  &  $\asmz$ & for default $\asmz$ & \\
\hline
HERAPDFv1.0  & 0.1176  & 95.1  & 81.7 \\
CT10         & 0.1180  & 94.5  & 88.2 \\
ABKM09NLO    & 0.1179  & 76.5  & 76.5 \\ 
NNPDFv2.1    & 0.1190  & 60.5  & 59.9 \\
MSTW2008NLO  & 0.1202  & 59.5  & 59.5 \\
\end{tabular}
\end{ruledtabular}
\end{table}

In summary, we have presented the first measurement of the 
inclusive three-jet differential cross section 
as a function of $\Mtj$ in $\ppbar$ collisions 
at a center of mass energy of $\sqrt{s}=1.96\, \mathrm{TeV}$. 
The three-jet cross section is measured 
in five scenarios, in different rapidity regions and for different 
requirements for the jet transverse momenta.
The data are compared to pQCD calculations in next-to-leading order 
in the strong coupling constant for different PDF parametrizations,
by computing $\chi^2$ values for different scale choices and 
different $\asmz$ values.
The best description of the data is obtained for the MSTW2008NLO 
and NNPDFv2.1 PDF parametrizations which describe both 
the normalization and the shape of the observed $\Mtj$ spectra. 
The PDF parametrizations from ABKM09NLO give a reasonable description
of the data, although with a slightly different shape of the $\Mtj$ spectrum.
The central results from the 
CT10 and HERAPDFv1.0 PDF sets predict a different $\Mtj$ shape
and are in poorer agreement with the data.

\input{acknowledgement}
\end{document}

%% file: author_list.tex
\affiliation{Universidad de Buenos Aires, Buenos Aires, Argentina}
\affiliation{LAFEX, Centro Brasileiro de Pesquisas F{\'\i}sicas, Rio de Janeiro, Brazil}
\affiliation{Universidade do Estado do Rio de Janeiro, Rio de Janeiro, Brazil}
\affiliation{Universidade Federal do ABC, Santo Andr\'e, Brazil}
\affiliation{Instituto de F\'{\i}sica Te\'orica, Universidade Estadual Paulista, S\~ao Paulo, Brazil}
\affiliation{Simon Fraser University, Vancouver, British Columbia, and York University, Toronto, Ontario, Canada}
\affiliation{University of Science and Technology of China, Hefei, People's Republic of China}
\affiliation{Universidad de los Andes, Bogot\'{a}, Colombia}
\affiliation{Charles University, Faculty of Mathematics and Physics, Center for Particle Physics, Prague, Czech Republic}
\affiliation{Czech Technical University in Prague, Prague, Czech Republic}
\affiliation{Center for Particle Physics, Institute of Physics, Academy of Sciences of the Czech Republic, Prague, Czech Republic}
\affiliation{Universidad San Francisco de Quito, Quito, Ecuador}
\affiliation{LPC, Universit\'e Blaise Pascal, CNRS/IN2P3, Clermont, France}
\affiliation{LPSC, Universit\'e Joseph Fourier Grenoble 1, CNRS/IN2P3, Institut National Polytechnique de Grenoble, Grenoble, France}
\affiliation{CPPM, Aix-Marseille Universit\'e, CNRS/IN2P3, Marseille, France}
\affiliation{LAL, Universit\'e Paris-Sud, CNRS/IN2P3, Orsay, France}
\affiliation{LPNHE, Universit\'es Paris VI and VII, CNRS/IN2P3, Paris, France}
\affiliation{CEA, Irfu, SPP, Saclay, France}
\affiliation{IPHC, Universit\'e de Strasbourg, CNRS/IN2P3, Strasbourg, France}
\affiliation{IPNL, Universit\'e Lyon 1, CNRS/IN2P3, Villeurbanne, France and Universit\'e de Lyon, Lyon, France}
\affiliation{III. Physikalisches Institut A, RWTH Aachen University, Aachen, Germany}
\affiliation{Physikalisches Institut, Universit{\"a}t Freiburg, Freiburg, Germany}
\affiliation{II. Physikalisches Institut, Georg-August-Universit{\"a}t G\"ottingen, G\"ottingen, Germany}
\affiliation{Institut f{\"u}r Physik, Universit{\"a}t Mainz, Mainz, Germany}
\affiliation{Ludwig-Maximilians-Universit{\"a}t M{\"u}nchen, M{\"u}nchen, Germany}
\affiliation{Fachbereich Physik, Bergische Universit{\"a}t Wuppertal, Wuppertal, Germany}
\affiliation{Panjab University, Chandigarh, India}
\affiliation{Delhi University, Delhi, India}
\affiliation{Tata Institute of Fundamental Research, Mumbai, India}
\affiliation{University College Dublin, Dublin, Ireland}
\affiliation{Korea Detector Laboratory, Korea University, Seoul, Korea}
\affiliation{CINVESTAV, Mexico City, Mexico}
\affiliation{FOM-Institute NIKHEF and University of Amsterdam/NIKHEF, Amsterdam, The Netherlands}
\affiliation{Radboud University Nijmegen/NIKHEF, Nijmegen, The Netherlands}
\affiliation{Joint Institute for Nuclear Research, Dubna, Russia}
\affiliation{Institute for Theoretical and Experimental Physics, Moscow, Russia}
\affiliation{Moscow State University, Moscow, Russia}
\affiliation{Institute for High Energy Physics, Protvino, Russia}
\affiliation{Petersburg Nuclear Physics Institute, St. Petersburg, Russia}
\affiliation{Instituci\'{o} Catalana de Recerca i Estudis Avan\c{c}ats (ICREA) and Institut de F\'{i}sica d'Altes Energies (IFAE), Barcelona, Spain}
\affiliation{Stockholm University, Stockholm and Uppsala University, Uppsala, Sweden}
\affiliation{Lancaster University, Lancaster LA1 4YB, United Kingdom}
\affiliation{Imperial College London, London SW7 2AZ, United Kingdom}
\affiliation{The University of Manchester, Manchester M13 9PL, United Kingdom}
\affiliation{University of Arizona, Tucson, Arizona 85721, USA}
\affiliation{University of California Riverside, Riverside, California 92521, USA}
\affiliation{Florida State University, Tallahassee, Florida 32306, USA}
\affiliation{Fermi National Accelerator Laboratory, Batavia, Illinois 60510, USA}
\affiliation{University of Illinois at Chicago, Chicago, Illinois 60607, USA}
\affiliation{Northern Illinois University, DeKalb, Illinois 60115, USA}
\affiliation{Northwestern University, Evanston, Illinois 60208, USA}
\affiliation{Indiana University, Bloomington, Indiana 47405, USA}
\affiliation{Purdue University Calumet, Hammond, Indiana 46323, USA}
\affiliation{University of Notre Dame, Notre Dame, Indiana 46556, USA}
\affiliation{Iowa State University, Ames, Iowa 50011, USA}
\affiliation{University of Kansas, Lawrence, Kansas 66045, USA}
\affiliation{Kansas State University, Manhattan, Kansas 66506, USA}
\affiliation{Louisiana Tech University, Ruston, Louisiana 71272, USA}
\affiliation{Boston University, Boston, Massachusetts 02215, USA}
\affiliation{Northeastern University, Boston, Massachusetts 02115, USA}
\affiliation{University of Michigan, Ann Arbor, Michigan 48109, USA}
\affiliation{Michigan State University, East Lansing, Michigan 48824, USA}
\affiliation{University of Mississippi, University, Mississippi 38677, USA}
\affiliation{University of Nebraska, Lincoln, Nebraska 68588, USA}
\affiliation{Rutgers University, Piscataway, New Jersey 08855, USA}
\affiliation{Princeton University, Princeton, New Jersey 08544, USA}
\affiliation{State University of New York, Buffalo, New York 14260, USA}
\affiliation{Columbia University, New York, New York 10027, USA}
\affiliation{University of Rochester, Rochester, New York 14627, USA}
\affiliation{State University of New York, Stony Brook, New York 11794, USA}
\affiliation{Brookhaven National Laboratory, Upton, New York 11973, USA}
\affiliation{Langston University, Langston, Oklahoma 73050, USA}
\affiliation{University of Oklahoma, Norman, Oklahoma 73019, USA}
\affiliation{Oklahoma State University, Stillwater, Oklahoma 74078, USA}
\affiliation{Brown University, Providence, Rhode Island 02912, USA}
\affiliation{University of Texas, Arlington, Texas 76019, USA}
\affiliation{Southern Methodist University, Dallas, Texas 75275, USA}
\affiliation{Rice University, Houston, Texas 77005, USA}
\affiliation{University of Virginia, Charlottesville, Virginia 22901, USA}
\affiliation{University of Washington, Seattle, Washington 98195, USA}
\author{V.M.~Abazov} \affiliation{Joint Institute for Nuclear Research, Dubna, Russia}
\author{B.~Abbott} \affiliation{University of Oklahoma, Norman, Oklahoma 73019, USA}
\author{B.S.~Acharya} \affiliation{Tata Institute of Fundamental Research, Mumbai, India}
\author{M.~Adams} \affiliation{University of Illinois at Chicago, Chicago, Illinois 60607, USA}
\author{T.~Adams} \affiliation{Florida State University, Tallahassee, Florida 32306, USA}
\author{G.D.~Alexeev} \affiliation{Joint Institute for Nuclear Research, Dubna, Russia}
\author{G.~Alkhazov} \affiliation{Petersburg Nuclear Physics Institute, St. Petersburg, Russia}
\author{A.~Alton$^{a}$} \affiliation{University of Michigan, Ann Arbor, Michigan 48109, USA}
\author{G.~Alverson} \affiliation{Northeastern University, Boston, Massachusetts 02115, USA}
\author{G.A.~Alves} \affiliation{LAFEX, Centro Brasileiro de Pesquisas F{\'\i}sicas, Rio de Janeiro, Brazil}
% ...
%
%
\collaboration{The D0 Collaboration\footnote{with visitors from
$^{a}$Augustana College, Sioux Falls, SD, USA,
$^{b}$The University of Liverpool, Liverpool, UK,
$^{c}$SLAC, Menlo Park, CA, USA,
$^{d}$University College London, London, UK,
$^{e}$Centro de Investigacion en Computacion - IPN, Mexico City, Mexico,
$^{f}$ECFM, Universidad Autonoma de Sinaloa, Culiac\'an, Mexico,
and 
$^{g}$Universit{\"a}t Bern, Bern, Switzerland.
}} \noaffiliation
\vskip 0.25cm

%% file: m3j-table.tex
\begin{table*}[!p]    

\caption{The three-jet differential cross section $d\sigma_{\text{3jet}} / d\Mtj$
and the theoretical predictions based on NLO pQCD (for MSTW200NLO PDFs 
with $\as(M_Z)=0.1202$)
plus non-perturbative corrections,
for renormalization and factorization scales 
$\mu_r = \mu_f = (p_{T1}+p_{T2}+p_{T3})/3$.
\label{tab:data}}
\begin{tabular*}{\textwidth}{@{\extracolsep{\fill}} lll  D{.}{.}{2.6} ll 
rrr}
\hline
Mass & Central & Measured & 
\multicolumn{1}{l}{Statistical} & 
Systematic & Theory & 
\multicolumn{3}{l}{Non-perturbative corrections} \\
\cline{7-9}
range & mass & cross section & 
\multicolumn{1}{l}{uncertainty} & 
uncertainty & cross section & Hadronization &
 Underlying & Total\\
TeV & TeV & pb/TeV & 
 \multicolumn{1}{l}{\%} & 
\% & pb/TeV & \% &
 event \,\, \%   & \% \\
\hline 
\multicolumn{9}{c}{for $|y|<0.8$ and $p_{T3}>40\,$GeV }\\ 
\hline 
0.40--0.45  & 0.424 & 1.93 $\times$ 10$^{2}$  & 1.8  & $+$11.2, $-$11.5 & 1.99 $\times$ 10$^{2}$ &  $-$9.0  &  9.1  & $-$0.7 \\
0.45--0.50  & 0.474 & 1.12 $\times$ 10$^{2}$  & 2.2  & $+$12.0, $-$11.1 & 1.17 $\times$ 10$^{2}$ &  $-$8.0  &  9.0  & $-$0.2 \\
0.50--0.55  & 0.524 & 5.89 $\times$ 10$^{1}$  & 2.9  & $+$12.2, $-$12.1 & 6.08 $\times$ 10$^{1}$ &  $-$7.5  &  8.7  &  0.6\\
0.55--0.61  & 0.578 & 2.76 $\times$ 10$^{1}$  & 3.8  & $+$11.6, $-$12.0 & 2.97 $\times$ 10$^{1}$ &  $-$7.1  &  8.2  &  0.5 \\
0.61--0.67  & 0.638 & 1.22 $\times$ 10$^{1}$  & 5.6  & $+$12.5, $-$11.8 & 1.33 $\times$ 10$^{1}$ &  $-$6.9  &  8.0  &  0.6\\
0.67--0.74  & 0.703 & 5.75 $\times$ 10$^{0}$  & 6.4  & $+$12.4, $-$13.5 & 5.98 $\times$ 10$^{0}$ &  $-$6.7  &  8.0  &  0.7\\
0.74--0.81  & 0.773 & 2.56 $\times$ 10$^{0}$  & 9.6  & $+$14.5, $-$13.6 & 2.45 $\times$ 10$^{0}$ &  $-$6.8  &  7.5  &  0.2\\
0.81--0.90  & 0.851 & 8.47 $\times$ 10$^{-1}$ & 14.8 & $+$14.9, $-$14.3 & 8.99 $\times$ 10$^{-1}$ &  $-$6.9  &  6.3  & $-$1.1\\
0.90--1.10  & 0.976 & 1.85 $\times$ 10$^{-1}$ & 21.3 & $+$18.1, $-$17.2 & 1.64 $\times$ 10$^{-1}$ &  $-$7.0  &  5.3  & $-$2.1\\
\hline 
\multicolumn{9}{c}{for $|y|<1.6$ and $p_{T3}>40\,$GeV }\\ 
\hline 
0.40--0.45  & 0.434 & 1.01 $\times$ 10$^{3}$  & 0.8  & $+$11.7, $-$12.3 & 1.10 $\times$ 10$^{3}$  & $-$10.0 &  8.4 & $-$2.4 \\
0.45--0.50  & 0.476 & 8.74 $\times$ 10$^{2}$  & 0.9  & $+$13.3, $-$12.5 & 9.95 $\times$ 10$^{2}$  &  $-$8.9 &  9.9 &  0.2 \\
0.50--0.55  & 0.525 & 6.28 $\times$ 10$^{2}$  & 1.0  & $+$12.6, $-$13.0 & 7.18 $\times$ 10$^{2}$  &  $-$8.3 & 10.6 &  1.4 \\
0.55--0.61  & 0.579 & 3.95 $\times$ 10$^{2}$  & 1.1  & $+$14.3, $-$12.5 & 4.47 $\times$ 10$^{2}$  &  $-$8.2 & 10.7 &  1.7 \\
0.61--0.67  & 0.639 & 2.08 $\times$ 10$^{2}$  & 1.5  & $+$14.5, $-$13.9 & 2.42 $\times$ 10$^{2}$  &  $-$8.0 & 10.5 &  1.6 \\
0.67--0.74  & 0.703 & 1.00 $\times$ 10$^{2}$  & 1.9  & $+$14.7, $-$14.9 & 1.18 $\times$ 10$^{2}$  &  $-$7.8 & 10.1 &  1.5 \\
0.74--0.81  & 0.773 & 4.54 $\times$ 10$^{1}$  & 2.8  & $+$16.4, $-$15.9 & 5.14 $\times$ 10$^{1}$  &  $-$7.7 &  9.8 &  1.4 \\
0.81--0.90  & 0.851 & 1.64 $\times$ 10$^{1}$  & 4.0  & $+$17.2, $-$15.6 & 1.91 $\times$ 10$^{1}$  &  $-$7.8 &  9.8 &  1.2 \\
0.90--1.10  & 0.978 & 3.18 $\times$ 10$^{0}$  & 6.0  & $+$20.3, $-$20.8 & 3.47 $\times$ 10$^{0}$  &  $-$8.2 &  9.5 &  0.5 \\
1.10--1.50  & 1.215 & 8.71 $\times$ 10$^{-2}$ & 20.6 & $+$27.4, $-$26.5 & 1.03 $\times$ 10$^{-1}$  &  $-$9.1 & 8.4 & $-$1.5 \\
\hline 
\multicolumn{9}{c}{for $|y|<2.4$ and $p_{T3}>40\,$GeV }\\ 
\hline 
0.40--0.45  & 0.419 & 1.23 $\times$ 10$^{3}$  & 0.7 & $+$12.1, $-$12.5 & 1.33 $\times$ 10$^{3}$  & $-$10.5 &  8.6 & $-$2.8 \\
0.45--0.50  & 0.477 & 1.17 $\times$ 10$^{3}$  & 0.8 & $+$13.6, $-$12.5 & 1.30 $\times$ 10$^{3}$  &  $-$9.6 &  9.9 & $-$0.7 \\
0.50--0.55  & 0.526 & 9.23 $\times$ 10$^{2}$  & 0.8 & $+$13.0, $-$13.1 & 1.05 $\times$ 10$^{3}$  &  $-$9.2 & 10.5 &  0.3 \\
0.55--0.61  & 0.580 & 6.68 $\times$ 10$^{2}$  & 0.9 & $+$14.6, $-$13.3 & 7.47 $\times$ 10$^{2}$  &  $-$9.1 & 10.7 &  0.7 \\
0.61--0.67  & 0.639 & 4.21 $\times$ 10$^{2}$  & 1.1 & $+$14.9, $-$14.7 & 4.81 $\times$ 10$^{2}$  &  $-$9.1 & 10.8 &  0.8 \\
0.67--0.74  & 0.704 & 2.48 $\times$ 10$^{2}$  & 1.3 & $+$15.5, $-$15.8 & 2.81 $\times$ 10$^{2}$  &  $-$9.2 & 11.0 &  0.8 \\
0.74--0.81  & 0.773 & 1.33 $\times$ 10$^{2}$  & 1.7 & $+$17.3, $-$15.6 & 1.50 $\times$ 10$^{2}$  &  $-$9.4 & 11.6 &  1.0 \\
0.81--0.90  & 0.852 & 6.01 $\times$ 10$^{1}$  & 2.3 & $+$18.1, $-$18.9 & 6.96 $\times$ 10$^{1}$  & $-$9.8 & 12.3 &  1.3 \\
0.90--1.10  & 0.983 & 1.38 $\times$ 10$^{1}$  & 3.1 & $+$20.6, $-$21.2 & 1.60 $\times$ 10$^{1}$  & $-$10.0 & 12.8 &  1.5 \\
1.10--1.50  & 1.225 & 5.40 $\times$ 10$^{-1}$ & 10.3 & $+$29.5, $-$27.2 & 6.54 $\times$ 10$^{-1}$  & $-$10.1 & 13.7 &  2.2 \\
\hline 
\multicolumn{9}{c}{for $|y|<2.4$ and $p_{T3}>70\,$GeV }\\ 
\hline 
0.40--0.45  & 0.421 & 3.84 $\times$ 10$^{2}$  & 1.4 & $+$12.2, $-$12.5 & 4.24 $\times$ 10$^{2}$  & $-$11.6 &  7.5 & $-$5.0 \\
0.45--0.50  & 0.481 & 4.29 $\times$ 10$^{2}$  & 1.3 & $+$13.0, $-$12.7 & 4.91 $\times$ 10$^{2}$  & $-$10.1 &  9.2 & $-$1.8 \\
0.50--0.55  & 0.526 & 3.71 $\times$ 10$^{2}$  & 1.3 & $+$13.2, $-$13.6 & 4.29 $\times$ 10$^{2}$  &  $-$9.5 &  10.0 & $-$0.5 \\
0.55--0.61  & 0.580 & 2.87 $\times$ 10$^{2}$  & 1.3 & $+$13.7, $-$13.5 & 3.29 $\times$ 10$^{2}$  &  $-$9.5 &  10.1 & $-$0.3 \\
0.61--0.67  & 0.639 & 1.90 $\times$ 10$^{2}$  & 1.6 & $+$14.8, $-$14.0 & 2.22 $\times$ 10$^{2}$  &  $-$9.5 &  10.1 & $-$0.3 \\
0.67--0.74  & 0.704 & 1.20 $\times$ 10$^{2}$  & 1.9 & $+$15.6, $-$15.5 & 1.38 $\times$ 10$^{2}$  &  $-$9.5 & 10.3 & $-$0.2 \\
0.74--0.81  & 0.774 & 6.52 $\times$ 10$^{1}$  & 2.5 & $+$16.7, $-$14.8 & 7.68 $\times$ 10$^{1}$  &  $-$9.6 & 10.8 &  0.2 \\
0.81--0.90  & 0.853 & 3.18 $\times$ 10$^{1}$  & 3.1 & $+$17.6, $-$19.6 & 3.80 $\times$ 10$^{1}$  &  $-$9.9 & 11.6 &  0.6 \\
0.90--1.10  & 0.985 & 8.27 $\times$ 10$^{0}$  & 4.0 & $+$21.5, $-$20.7 & 9.71 $\times$ 10$^{0}$  &  $-$10.2 & 12.1 &  0.7 \\
1.10--1.50  & 1.235 & 3.40 $\times$ 10$^{-1}$ & 13.0& $+$28.6, $-$24.5 & 4.21 $\times$ 10$^{-1}$  &  $-$10.3 & 12.7 &  1.1 \\
\hline 
\multicolumn{9}{c}{for $|y|<2.4$ and $p_{T3}>100\,$GeV }\\ 
\hline 
0.40--0.45  & 0.424 & 3.80 $\times$ 10$^{1}$  & 4.0  & $+$13.9, $-$13.5 & 4.80 $\times$ 10$^{1}$  & $-$13.8 &  5.3 & $-$9.3 \\
0.45--0.50  & 0.472 & 7.55 $\times$ 10$^{1}$  & 2.8  & $+$15.0, $-$11.9 & 8.72 $\times$ 10$^{1}$  & $-$11.2 &  6.6 & $-$5.3 \\
0.50--0.55  & 0.535 & 8.53 $\times$ 10$^{1}$  & 2.6  & $+$13.6, $-$13.7 & 9.61 $\times$ 10$^{1}$  &  $-$9.9 &  7.7 & $-$2.9 \\
0.55--0.61  & 0.581 & 7.56 $\times$ 10$^{1}$  & 2.4  & $+$12.9, $-$13.7 & 8.66 $\times$ 10$^{1}$  &  $-$9.5 &  8.6 & $-$1.7 \\
0.61--0.67  & 0.640 & 5.59 $\times$ 10$^{1}$  & 2.8  & $+$13.9, $-$13.1 & 6.60 $\times$ 10$^{1}$  &  $-$9.2 &  9.3 & $-$0.7 \\
0.67--0.74  & 0.704 & 3.84 $\times$ 10$^{1}$  & 3.1  & $+$15.0, $-$14.7 & 4.46 $\times$ 10$^{1}$  &  $-$8.9 &  9.8 & 0.1 \\
0.74--0.81  & 0.774 & 2.18 $\times$ 10$^{1}$  & 4.0  & $+$18.9, $-$16.4 & 2.69 $\times$ 10$^{1}$  &  $-$8.8 & 10.0 &  0.3 \\
0.81--0.90  & 0.853 & 1.16 $\times$ 10$^{1}$  & 4.8  & $+$16.5, $-$18.6 & 1.41 $\times$ 10$^{1}$  &  $-$9.3 & 10.1 &  $-$0.1 \\
0.90--1.10  & 0.985 & 3.38 $\times$ 10$^{0}$  & 6.0  & $+$22.2, $-$20.7 & 3.89 $\times$ 10$^{0}$  & $-$9.8 & 10.4 &  $-$0.5 \\
1.10--1.50  & 1.225 & 1.58 $\times$ 10$^{-1}$ & 18.4 & $+$25.0, $-$22.2 & 2.10 $\times$ 10$^{-1}$  & $-$10.1 & 11.5 & 0.3 \\
\hline
\end{tabular*}
\end{table*}

%% file: acknowledgement.tex
%
We thank the staffs at Fermilab and collaborating institutions,
and acknowledge support from the
DOE and NSF (USA);
CEA and CNRS/IN2P3 (France);
FASI, Rosatom and RFBR (Russia);
CNPq, FAPERJ, FAPESP and FUNDUNESP (Brazil);
DAE and DST (India);
Colciencias (Colombia);
CONACyT (Mexico);
KRF and KOSEF (Korea);
CONICET and UBACyT (Argentina);
FOM (The Netherlands);
STFC and the Royal Society (United Kingdom);
MSMT and GACR (Czech Republic);
CRC Program and NSERC (Canada);
BMBF and DFG (Germany);
SFI (Ireland);
The Swedish Research Council (Sweden);
and
CAS and CNSF (China).